\documentclass[final,3p,times,twocolumn]{elsarticle}

\usepackage{amssymb}
\usepackage{amsmath}
\usepackage{xcolor}
\usepackage{graphicx} 
\usepackage{caption}   
\usepackage{hyperref}
\usepackage{booktabs}
\usepackage{siunitx}
\usepackage{float}
\usepackage{placeins}
\usepackage{listings}
\usepackage{bm}
\usepackage{enumitem}

\lstdefinestyle{pythoncode}{
    language=Python,
    basicstyle=\ttfamily\scriptsize,
    keywordstyle=\color{blue}\bfseries,
    commentstyle=\color{gray}\itshape,
    stringstyle=\color{red},
    showstringspaces=false,
    breaklines=true,
    numbers=left,
    columns=fullflexible,
    numberstyle=\tiny\color{gray},
    numbersep=5pt,
    frame=tb,
    framexleftmargin=15pt,
    xleftmargin=15pt,
    tabsize=4,
    captionpos=b,
    morekeywords={as, with, yield, lambda, self, True, False, None},
    emphstyle=\color{green}\bfseries,
}

\hypersetup{
    colorlinks=true,
    linkcolor=blue,
    filecolor=magenta,      
    urlcolor=cyan,
    pdftitle={Overleaf Example},
    pdfpagemode=FullScreen,
    }

\journal{TBC}

\newcommand{\pyvale}{\textsc{Pyvale}}

\begin{document}

\begin{frontmatter}



\title{\pyvale: A Fast, Scalable, Open-Source 2D Digital Image Correlation (DIC) Engine Capable of Handling Gigapixel Images}


\author[ukaea]{Joel Hirst}
\author[ukaea,shf]{Lorna Sibson}
\author[ukaea]{Adel Tayeb}
\author[ukaea]{Ben Poole}
\author[ukaea]{Megan Sampson}
\author[ukaea]{Wiera Bielajewa}
\author[ukaea]{Michael Atkinson}
\author[ukaea]{Alex Marsh}
\author[ukaea]{Rory Spencer}
\author[ukaea]{Robert Hamill}
\author[ukaea]{Cory Hamelin}
\author[ukaea]{Allan Harte}
\author[ukaea]{Lloyd Fletcher}

\affiliation[ukaea]{organization={United Kingdom Atomic Energy Authority},
            addressline={Culham Campus}, 
            city={Abingdon},
            postcode={OX14 3DB}, 
            state={Oxfordshire},
            country={United Kingdom}}

\affiliation[shf]{organization={School of Electrical and Electronic Engineering, University of Sheffield},
            addressline={Mappin Street}, 
            city={Sheffield},
            postcode={S1 4DT }, 
            country={United Kingdom}}

\begin{abstract}
\textbf{Background:}
Digital Image Correlation (DIC) is a widely used full‑field measurement technique, but both open‑source and commercial packages often have limitations such as operating‑system restrictions, lack of support for deployment on computing clusters, and poor scalability to gigapixel‑scale images common in Scanning Electron Microscopy DIC (SEM-DIC).
\textbf{Objective:}
\pyvale\ is an open‑source software package designed for sensor simulation, uncertainty quantification, placement optimization, and calibration/validation. A key component of this is the development of a dedicated 2D DIC module intended for standalone use and integration within broader workflows.
\textbf{Methods:}
\pyvale\ provides a user‑friendly Python interface with performant compiled routines underneath. At its core is a multithreaded, reliability‑guided DIC algorithm. Its open‑source MIT license enables wide deployment, including on computing clusters and in automated pipelines.
\textbf{Results:}
Benchmarking with the publicly available 2D DIC challenge 2.0 dataset shows that \pyvale\ achieves metrological performance comparable to existing commercial and open-source DIC codes. It can correlate gigapixel‑scale image pairs in under 5 minutes on high-specification desktop workstations, with memory peaking at approximately 50 GB.
\textbf{Conclusions:}
\pyvale's strong metrological foundation, coupled with its scalability for SEM‑DIC, positions it as a platform for sustained, community‑driven development. Its design and licensing provide a foundation for future improvements in open‑source DIC and integration into experimental design and validation workflows.
\end{abstract}



\end{frontmatter}

\section{Introduction}
\label{intro}
Digital Image Correlation (DIC) is a widely used technique for measuring shape deformation and motion \cite{schreier2009image, Pan2018, IDICS2018}. The idea is to obtain displacements by correlating gray-level values between a reference image and one or more images captured post deformation. In local subset DIC, a non-linear optimization routine is used to calculate the shape function (rigid, affine, quadratic etc.) parameters that describe the displacement and deformation of the subset between reference and deformed images. A single-camera setup (2D-DIC) enables measurement of in-plane displacements only, while stereo configurations extend this to capture out-of-plane motion. Since the first use of DIC in experimental mechanics by Peters and Ranson in 1982 \cite{peters1982digital}, the field has seen vast improvements in the underlying computational and numerical algorithms. Some key algorithmic developments include the use of the Newton-Raphson (NR) method for minimizing cost functions \cite{chu1985applications}, using fast-Fourier transforms (FFTs) for integer pixel displacement estimation \cite{chen1993digital}, Reliability-Guided DIC (RG-DIC) \cite{rg}, and understanding systematic errors due to image interpolation at the gray-level \cite{schreier2000systematic}.
\\
\\
These methodological advancements have, in turn, led to the creation of several well-established software packages. Commercial codes include MatchID \cite{matchid}, DaVis \cite{lavision_davis}, VIC-2D/VIC-3D \cite{correlatedsolutions2025} to name a few, while open-source local subset DIC codes, such as Ncorr \cite{Blaber2015}, OpenCorr \cite{jiang2023opencorr}, and DICe \cite{dice}, are widely used in academic and research settings. Ncorr comes as a MATLAB extension with an easy installation process and intuitive interactive Graphical User Interface (GUI) for correlation configuration and Region of Interest (ROI) selection. However, there is very little scripting support for the extension, making systematic studies with correlation parameters a challenge. OpenCorr and DICe are both C++ codes that provide prebuilt executables for Windows. For Linux operating systems, the package and dependencies must be built from source which makes installation a time consuming process. Another notable mention goes to $\mu$DIC  \cite{olufsen2020mudic}, which is a well documented, open-source global DIC code. $\mu$DIC is written purely in Python which, while making usage straightforward, means it struggles computationally with higher resolution images.
\\
\\
Commercial tools are built with user friendliness in mind, offering intuitive GUIs and reliable performance. However, they often come with significant financial cost and limited flexibility, with users typically unable to access or modify the underlying algorithms beyond what the vendor is willing to implement or permit. Another key limitation of commercial codes is licensing that restricts the ability of users to perform large distributed array jobs on computing clusters for experimental design and uncertainty quantification. Open-source alternatives offer greater adaptability, allowing users to modify and extend the software to suit specific needs. These open-source codes also have the benefit that they can be run without licensing restrictions on computing clusters. However, this flexibility often comes at the expense of usability or performance. DIC codes written in commonly used data analysis languages, such as Python, can be user-friendly but struggle with memory usage and computational performance. In contrast, lower-level compiled languages (e.g., C/C++) offer improved performance and memory management but require more complex setup and dependency management, beyond the typical capabilities of engineers not focused on computational work. Furthermore, open-source tools that rely on proprietary environments such as MATLAB still require a paid license, which hinders accessibility. Therefore, there is a gap in available open-source DIC software that combines an intuitive and easy-to-use interface with the computational performance of compiled code through Python C extensions.
\\
\\
To study material deformation at the microscopic scale, techniques that exceed the resolution limits of optical microscopy ($\sim$200 nm) are essential. However, a major limitation of current open-source DIC tools is their inability to efficiently handle extremely large images, such as gigapixel-scale datasets required to capture representative areas with displacement resolution to reveal micro-mechanics modes of deformation from Scanning Electron Microscope (SEM) DIC. Electron-based imaging enables analysis at much smaller scales due to the shorter wavelength of electrons. SEM-DIC, particularly using field emission SEMs (FESEM), has gained traction in recent years \cite{lockwood1999use, kammers2013digital, lindfeldt2014using, liu2024digital, kammers2011small, jin2008micro, mello2017distortion}, though recent work explores more cost effective systems that use tungsten or LaB$_6$ emitters \cite{sem_LaB6}. A key challenge with SEM-DIC is the speckle pattern generation. As per the DIC good practices guide, the ideal size of each speckle should be between 3-5 pixels \cite{IDICS2018}. At SEM resolutions this translates to a physical speckle size in the nanometer range. An in-depth review of methods for speckle pattern generation at these scales can be found in Ref. \cite{kammers2011small}. The fine spatial resolution combined with a large ROI can still produce comparatively large subset displacements in pixel units. It is often the case that with an optimal speckle pattern size of 3-5 pixels, subset displacements can be well in excess of 1000 pixels. To capture these large deformations while maintaining sufficiently high spatial resolution, SEM captures are stitched together. With images approaching gigapixel sizes, memory management and processing speed quickly become one of the most important factors when selecting a DIC code. 
\\
\\
To address this, we have designed the 2D DIC module in \pyvale\ to meet the following criteria:
(i) open-source with permissive licensing allowing scalable usage for large array jobs on the super-computing clusters,
(ii) a simple and accessible Python based user interface,
(iii) memory efficient, parallelized low-level code for good computational performance (written in C/C++),
(iv) straightforward installation across all major operating systems and computing architectures, and
(v) the ability to process gigapixel size images for SEM-DIC applications.
This gives \pyvale\ users the simplicity of a Python interface combined with the performance of optimized compiled code, all under a free and open-source license that supports deployment on computing clusters.
\\
\\
Here we will present an in-depth overview of \pyvale's user interface, the algorithms used, as well as benchmarking current metrological and computational performance. The paper is organized as follows: Section \ref{sec:design} presents the design and implementation of \pyvale, including the user interface and key algorithms. Section \ref{sec:DIC2} provides numerical validation against the 2D DIC challenge. Section \ref{sec:SEM} evaluates performance on gigapixel SEM-DIC images, with a focus on runtime and memory scaling. Section \ref{sec:runtime_comparison} presents a runtime comparison to other open-source codes using both SEM-DIC and synthetic speckle pattern images. Finally, section \ref{sec:future} provides an outlook on future DIC development and user engagement.
\begin{figure}[t!]
  \centering
  \includegraphics[width=1.0\linewidth]{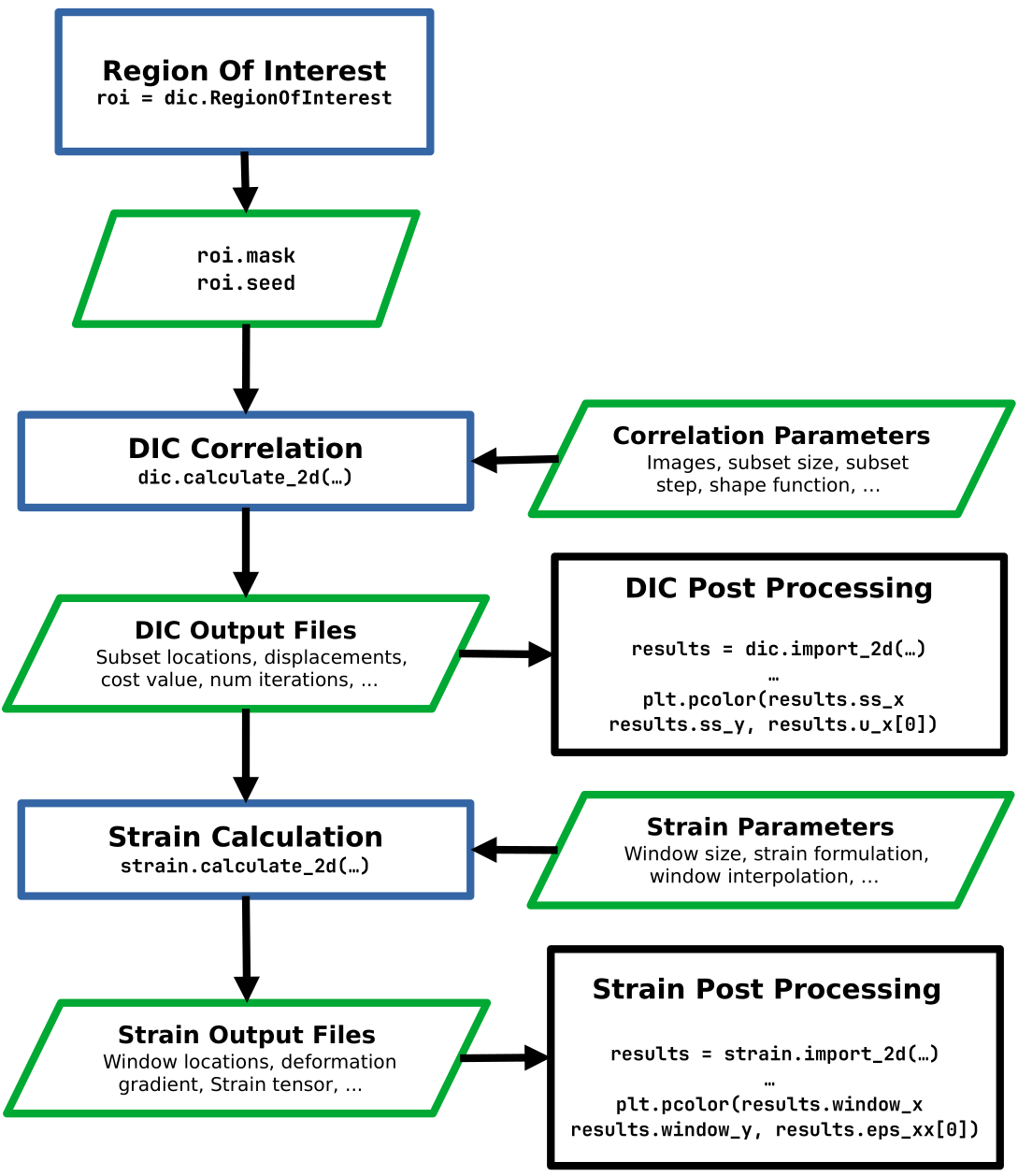}
  \caption{High-level flowchart illustrating the \pyvale\ workflow. Each box shows the required core functions, parameters, and arguments. The two post-processing boxes give examples of how data can be plotted with \texttt{matplotlib}, although the data can also be imported into other analysis packages (e.g., Excel or Origin) at user discretion.}
  \label{fig:workflow}
\end{figure}
\section{Design \& Implementation}
\label{sec:design}
\subsection{User Interface \& Workflow}
At a high level, \pyvale's DIC engine is a Python module that uses Python C extensions to make calls down to lower level C/C++ code for computationally intensive calculations. When installed via Python’s \texttt{pip} package manager on major operating systems (Windows 10/11, macOS, Linux), the C/C++ components are precompiled which significantly simplifies the setup process. Typical users are not required to configure or install multiple dependencies manually; instead, they receive a ready-to-use package for their system. For users needing greater control or wishing to make modifications, \pyvale\ also supports manual compilation from source.
\\
\\
A flowchart summarizing key components of the \pyvale\ DIC workflow from a user’s perspective is shown in Fig. \ref{fig:workflow}. To run \pyvale's DIC engine with the minimum required amount of input, the user must specify the ROI over which they want to perform the correlation and some fundamental correlation parameters such as subset-size and subset-step.
The ROI selection can be completed in one of three ways: (i) interactively via a built-in GUI tool, (ii) by defining a configuration in a ``.yaml'' file, or (iii) by creating a boolean array either manually (for example using standard \texttt{NumPy} array operations) or using the helper functions included in \pyvale. The ability to programmatically manipulate the ROI without having to open images is particularly useful for larger resolutions (i.e. gigapixel-scale), which can cause lag/crashes when opened within a GUI.
\\
\\
Once an ROI has been chosen, the user can proceed with a numerical correlation to obtain subset displacement values. A simple DIC calculation example can be found in Listing \ref{list:dic}. Here, the user sets the reference and deformed images prior to calling the interactive ROI selection tool. The DIC analysis is then performed using a subset-size of 31 pixels and step of 5 pixels, with the ROI mask and starting seed location selected during the interactive ROI session. For each deformed image, there will be a corresponding result file containing correlation data. There are a wide range of additional arguments that can be passed to the DIC engine, allowing the user to tailor the setup to suit their needs. These include the ability to manipulate the convergence conditions, output file formatting, number of threads. Further details can be found in the package documentation \cite{pyvale_docs}.
\begin{lstlisting}[style=pythoncode,  float=t!, label={list:dic}, caption=A simple example using the interactive ROI selection and 2D DIC in \pyvale. Accurate as of \pyvale\ release 2026.1.1.]
import pyvale.dic as dic

ref_image="./ref_0000.tiff"
def_image="./def_*.tiff"

roi = dic.RegionOfInterest(ref_image)
roi.interactive_selection()

dic.calculate_2d(reference=ref_image,
                 deformed=def_image,
                 roi_mask=roi.mask,
                 seed=roi.seed,
                 subset_size=31,
                 subset_step=15,
                 output_basepath="./")
\end{lstlisting}
By default, results are saved in human-readable comma-separated value (CSV) format in the current working directory. Each file will be prefixed with ``\texttt{dic\_results\_}'' followed by the original name of the deformed image unless otherwise specified. The delimiter, output folder and filename prefix can be altered if required. Data can also be saved in a non-human-readable binary format, which is particularly useful for large images where writing to human-readable text format becomes a time-consuming part of the workflow. In both cases, \pyvale\ provides the ability to import results into a structured dataclass composed of \texttt{NumPy} arrays for custom analysis and visualization.
\\
\\
While \pyvale\ does not currently include built-in plotting functionality for displacement and strain data, many Python packages (e.g., Matplotlib \cite{matplotlib}, Plotly \cite{plotly}) can be used as part of the same script to visualize results. A simple example demonstrating how to import DIC data and generate a 2D plot of the vertical displacement in the first deformed image is shown in Listing \ref{list:import}.
\\
\\
\begin{lstlisting}[style=pythoncode,  float=t!, label={list:import}, caption=Example of importing \pyvale\ DIC displacement data into Python and plotting the vertical displacement field for deformed image 0. Accurate as of \pyvale\ release 2026.1.1.]
import matplotlib.pyplot as plt
import pyvale.dic as dic

dic_data = dic.import_2d(data="./dic_results_*",
                         binary=False,
                         delimiter=",")

# plot of vertical displacement for first deformation image.
plt.pcolor(dic_data.ss_x, 
           dic_data.ss_y, 
           dic_data.u_y[0]) # [image, y, x]
plt.show()
\end{lstlisting}
\pyvale\ DIC result files contain displacement fields and can be used to a perform a deformation gradient and strain field calculation. A user can pass the type of strain formulation to use, the strain window size, whether to use bilinear or biquadratic interpolation on the window, and options for output data file formatting. Strain data can be imported in a similar way to DIC data. An example showing a simple strain calculation and importing the results can be found in Listing \ref{list:strain}. 
\subsection{Underlying DIC Algorithms}
\label{sec:algo}
\begin{figure}[htbp]
  \centering
  \includegraphics[width=1.0\linewidth, trim=100 60 100 60, clip]{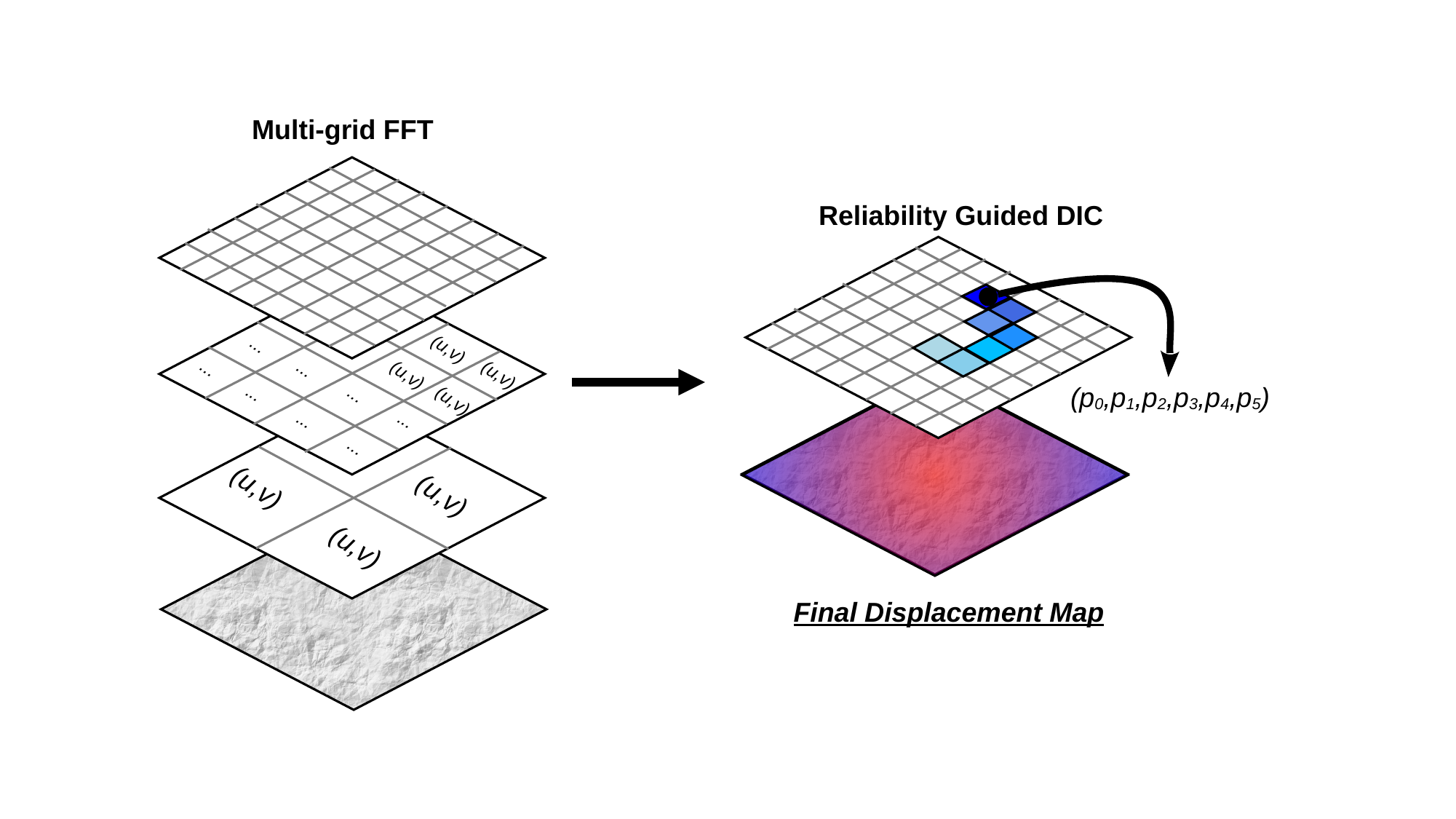}
  \caption{Schematic of the main algorithmic approach taken in Pyvale. The first stage is to perform a FFT cross correlation to obtain rigid displacement values for decreasing window size. The final window size is then used to seed the RG-DIC to obtain higher order shape function parameters if required.}
  \label{fig:algo_schematic}
\end{figure}

The capability of \pyvale\ to handle gigapixel-scale images and large displacement fields is attributed to a two-stage processing approach. Initially, a multi-window FFT based technique, which is a key method in Particle Image Velocimetry (PIV) \cite{scarano1999iterative, scarano2000advances, openpiv}, is applied to estimate rigid displacements for all subsets within the user-specified ROI. Similar methods have been used in previous versions of DaVis \cite{lavision_davis}. The DaVis implementation of the multi-window FFT method has previously been benchmarked against other commercial and open-source codes specifically for SEM-based micro-mechanics applications in \cite{lunt2020comparison}. The multi-window procedure begins by splitting images into square subsets initially with dimensions larger than the user estimated maximum displacement. Within each window, a normalized FFT-based cross-correlation (FFTCC) is computed to estimate the rigid-body displacement. The open-source PocketFFT library \cite{pocketfft} is used for all FFT calculations in \pyvale. The FFT-based normalized cross-correlation is given by:
\begin{equation}
R_{fg}(x,y) \;=\;
\mathcal{F}^{-1} \left\{ 
\frac{\mathcal{F}\{f(x,y)\}\,\overline{\mathcal{F}\{g(x,y)\}}}
{\big|\mathcal{F}\{f(x,y)\}\big|\,\big|\mathcal{F}\{g(x,y)\}\big|}
\right\},
\end{equation}
\begin{lstlisting}[style=pythoncode, float=t!, label={list:strain}, caption=An example of performing a strain calculation and plotting the resulting horizontal strain field. Accurate as of \pyvale\ release 2026.1.1.]
import matplotlib.pyplot as plt
import pyvale.dic as dic
import pyvale.strain as strain

strain.calculate_2d(data="./dic_results_*",
                    input_binary=False,
                    input_delimiter=",",
                    window_size=5,
                    window_element=9
                    output_basepath="./")

strain_data = strain.import_2d(data="strain_*",
                               delimiter=",")

# plot xx component of 2D strain tensor for 
# first deformed image.
plt.pcolor(strain_data.window_x, 
           strain_data.window_y, 
           strain_data.eps_xx[0]) # [image, y, x]

plt.show()
\end{lstlisting}
where $\mathcal{F}$ and $\mathcal{F}^{-1}$ denote the forward and inverse 2D Fourier transform respectively, $\overline{\mathcal{F}\{g(x,y)\}}$ is the complex conjugate of the Fourier transform of $g(x,y)$, where $f(x,y)$ is the reference window, and $g(x,y)$ is the corresponding window in the deformed image. The location of the peak in the cross-correlation map $R_{fg}(x,y)$ yields the rigid displacement vector $(u,v)$ for that particular window. After estimating the rigid shift, the window size is halved. The displacements, $(u_x,u_y)$, from the four nearest points in the previous, coarser window are then used as the initial guess for the rigid displacement in the newly halved window. This process of halving the window size continues iteratively until the user-defined subset size is reached. An illustration of this procedure can be found in the left-hand side of Fig \ref{fig:algo_schematic}. 
\\
\\
\pyvale\ determines window sizes automatically (excluding the user specified subset size) by selecting powers of two ($2^n$) descending from the next highest power of two above the user estimated maximum displacement. This exploits the inherently faster FFT computations for powers of two compared to arbitrary array lengths \cite{brigham1988fast}. As an example of this, take the case where a user estimates that the maximum displacement is approximately 800 pixels and specifies a subset-size of 17 pixels. The \pyvale\ algorithm will use window sizes of 1024, 512, 256, 128, 64, 32 and finally 17 to estimate the rigid displacement. The rigid values of the final window (in this case 17) will be used as seed values for the RG-DIC algorithm \cite{rg}. For micro-mechanics applications it may be desirable to only run the multi-window FFT correlation without a subsequent RG-DIC step as the contribution of the higher order shape function parameters can be small. In this case the user can set \texttt{method=`MULTIWINDOW'} when calling \texttt{dic.calculate\_2d}. To obtain subpixel displacement values using only the multi-window approach, a 2D gaussian is fitted in the vicinity of the correlation peak in Fourier space. Generally, it is recommended that the user use \texttt{method=`MULTIWINDOW\_RG'} (default), even for rigid displacements. A bicubic b-spline is used to obtain intensity values at subpixel locations.
\\
\\
For each subset, displacement is obtained by minimizing a cost function. The Sum of Squared Differences (SSD) compares gray-level intensities between the reference and deformed images:
\begin{equation}
\text{SSD} = \sum_i \big(f(x_{i},y_{i}) - g(x_{i},y_{i})\big)^{2}
\end{equation}
where $f(x_{i},y_{i})$ and $g(x_{i},y_{i})$ represent the gray-level intensity values at location $(x_{i},y_{i})$ in the reference and deformed images, respectively. To reduce sensitivity to intensity scaling, the Normalized SSD (NSSD) is defined as:
\begin{equation}
\text{NSSD} = \sum_i \left( \frac{f(x_{i},y_{i})}{\sqrt{\sum_{j} f(x_{j},y_{j})^{2}}} - \frac{g(x_{i},y_{i})}{\sqrt{\sum_{j} g(x_{j},y_{j})^{2}}} \right)^{2}
\end{equation}
The Zero-Normalized SSD (ZNSSD), used by default in \pyvale, further accounts for intensity offsets:
\begin{equation}
\text{ZNSSD} = \sum_i \left( \frac{\bar{f}(x_{i},y_{i})}{\sqrt{\sum_{j} \bar{f}(x_{j},y_{j})^{2}}} - \frac{\bar{g}(x_{i},y_{i})}{\sqrt{\sum_{j} \bar{g}(x_{j},y_{j})^{2}}} \right)^{2}
\end{equation}
where $\bar{f}(x_{i},y_{i}) = f(x_{i},y_{i}) - f_m$, and $f_m$ is the mean gray-level value of the subset. ZNSSD is invariant to both additive and multiplicative intensity changes. There are three options for shape functions that transforms pixel coordinates in the subset from the reference image to the deformed image:
\begin{equation}
\begin{split}
\xi(x_i,y_i, \mathbf{p}) =
\underbrace{\begin{bmatrix} p_0 \\ p_1 \end{bmatrix}}_{\text{rigid}}
+ \underbrace{\begin{bmatrix} 1+p_2 & p_3 \\ p_4 & 1+p_5 \end{bmatrix} 
\begin{bmatrix} x_i \\ y_i \end{bmatrix}}_{\text{affine}} \cdots \\
 \cdots + \underbrace{\begin{bmatrix} p_6 & p_7 & p_8 \\ p_9 & p_{10} & p_{11} \end{bmatrix} 
\begin{bmatrix} x_i^2 \\ x_iy_i \\ y_i^2 \end{bmatrix}}_{\text{quadratic}}
\end{split}
\end{equation}
Each higher-order shape function includes all terms from the lower-order functions: affine includes rigid terms, and quadratic includes both affine and rigid terms. 
\\
\\
For the non-linear optimization for each subset a Levenberg-Marquardt \cite{marquardt1963algorithm, levenberg1944method} algorithm is used. The convergence threshold and precision, as well as the maximum number of iterations for a given subset optimization can be altered by the user if needed. Default values are used when they are omitted by the user. There is a large amount of existing literature on optimizations of cost and shape functions in DIC and therefore the explanations are omitted here. Well explained mathematical descriptions can be found across various works \cite{schreier2009image, Blaber2015, Pan2018, pan2008study}.

\subsection{Strain Calculation and Formulations}
From the displacement fields, the in-plane strain components can be calculated using the spatial derivatives of the displacement field to capture local changes in geometry. The 2D deformation gradient matrix, $\mathbf{F}$, is given by the equation:
\begin{equation}
\mathbf{F} =
\begin{bmatrix}
1+\dfrac{\partial u_x}{\partial x} & \dfrac{\partial u_x}{\partial y} \\
\dfrac{\partial u_y}{\partial x} & 1+\dfrac{\partial u_y}{\partial y}
\end{bmatrix}
= \mathbf{I} + \nabla \boldsymbol{u},
\qquad
\boldsymbol{u}=
\begin{bmatrix}
u_x \\ u_y
\end{bmatrix}.
\end{equation}
To calculate the partial derivatives above, we calculate the gradient over a square window containing $N \times N$ displacement data points from the DIC calculation.  Because we are using displacement values from DIC, the strain window is dependent on the subset-step, $s$, and subset-size, $w$. These quantities, along with the size of the strain window, form what is typically referred to as the Virtual Strain Gauge, $VSG$:
\begin{equation}
VSG = (\textrm{N} - 1)s + w 
\end{equation}
Due to the noise in DIC measurements, smoothing is typically applied over the strain window. \pyvale\ supports bilinear and biquadratic smoothing over the strain window elements. The polynomial approximation is given by:
\begin{equation}
\boldsymbol{u}(x,y)=\mathbf{P}(x,y)\,\mathbf{c},
\end{equation}
where $\boldsymbol{u}=[\,u_x\;u_y\,]^T$, $\mathbf{P}(x,y)$ is a row vector of
basis terms, and $\mathbf{c}$ is a column vector of coefficients. The polynomial
basis is
\begin{equation}
\mathbf{P}(x,y)=
\begin{cases}
[1, x, y] & \text{bilinear},\\
[1, x, y, x^2, y^2, x^2y, xy^2, x^2y^2] & \text{biquadratic}
\end{cases}
\end{equation}
The coefficients are obtained by solving a linear least-squares problem, since the displacement field is
linearly parameterized in the unknown coefficients. Once the polynomial coefficients have been obtained the deformation gradient tensor and strain can be calculated. \pyvale\ supports the following strain tensor formulations:
\begin{itemize}[itemsep=1pt]
\item \textit{Green-Lagrange}: $\bm{\varepsilon}=\tfrac{1}{2}\left(\mathbf{F}^\mathsf{T}\mathbf{F}-\mathbf{I}\right)$
\item \textit{Hencky (logarithmic)}: $\bm{\varepsilon}=\ln(\sqrt{\mathbf{F}^\mathsf{T}\mathbf{F}})$
\item \textit{Euler-Almansi}: $\bm{\varepsilon}=\tfrac{1}{2}\left(\mathbf{I}-(\mathbf{F}\mathbf{F}^\mathsf{T})^{-1}\right)$
\item \textit{Biot (right / Lagrangian)}: $\bm{\varepsilon} = \sqrt{\mathbf{F}^\mathsf{T}\mathbf{F}}-\mathbf{I}$
\item \textit{Biot (left / Eulerian)}: $\bm{\varepsilon} =\sqrt{\mathbf{F}\mathbf{F}^\mathsf{T}}-\mathbf{I}$
\end{itemize}
where $\mathbf{I}$ the $2 \times 2$ identity matrix.
\begin{figure*}
  \centering
  \includegraphics[width=1.0\linewidth, trim=0 100 0 0, clip]{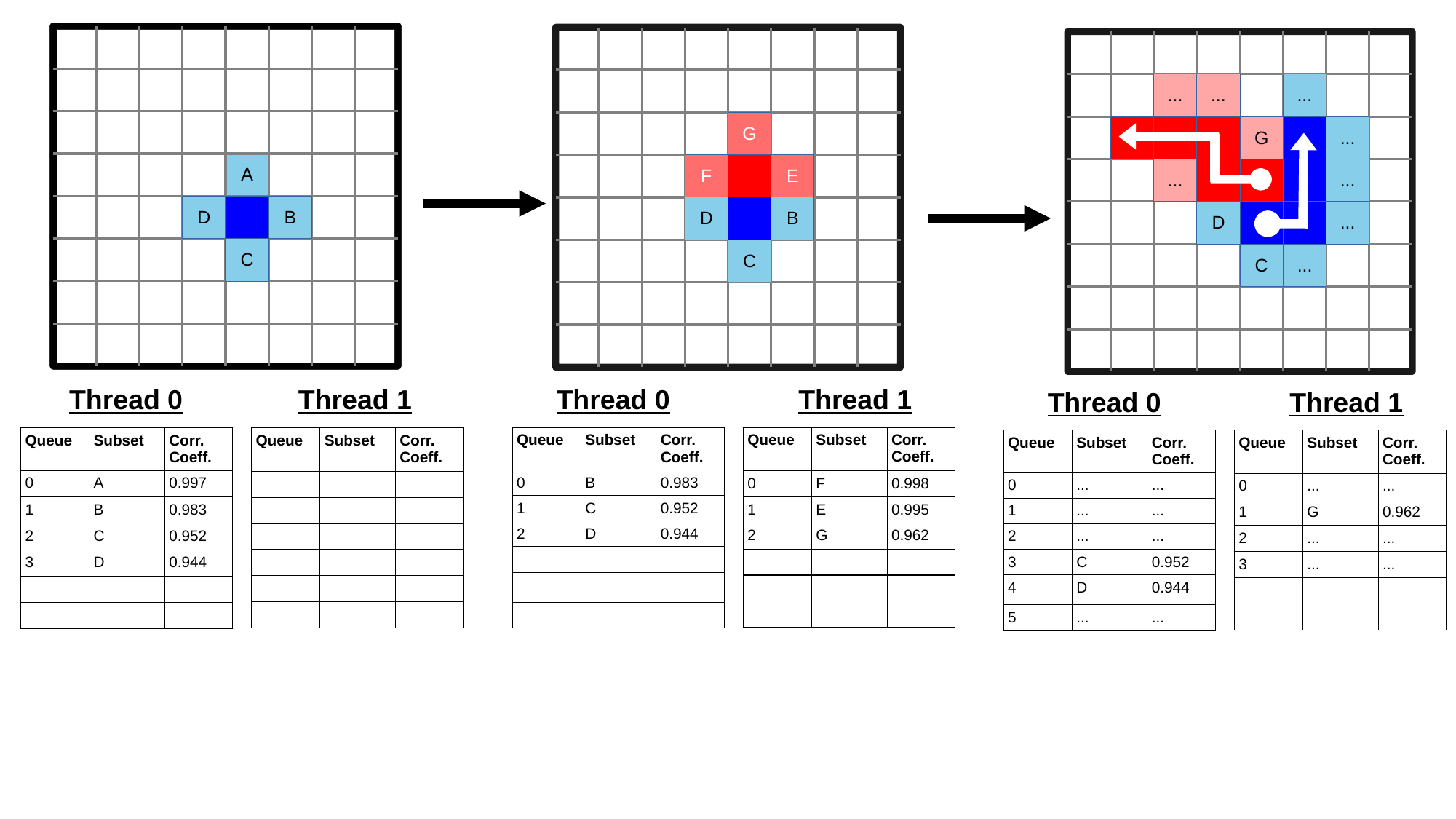}
  \caption{Schematic of the multithreadeded RG-DIC algorithm adopted in \pyvale\. Here it is shown for two threads. (Left) A single thread performs calculation for the seed point and its 4 neighbours. (Middle) The second thread then ``steals'' work from the queue of thread 0 to start its own RG-DIC prodecure and computes correlation values for the ``stolen''  point. These are then added to its own queue. (Right) A view of the queue sometime later where each thread has gone off in independent directions, note the position of subsets in the queue have changed depending on the ranking of the correlation coefficient.}
  \label{fig:rg_schematic}
\end{figure*}
\subsection{Parallelization}
\pyvale\ uses the Open Multi-Processing (OpenMP) library \cite{dagum1998openmp} to parallelize all major computations, including subset generation, bicubic interpolation setup, multi-window FFT-based rigid displacement estimation, and the RG-DIC algorithm. The mutli-window FFT-displacement estimation is parallelized by evenly distributing windows among threads, with each thread  storing rigid displacement values in a shared array to be used as initial estimates for subsequent iterations.
\\
\\
\pyvale\ uses a dynamic queuing parallelization strategy, which is contrasted against the thread-specific RG-DIC seed assignment approach used in the MATLAB extension NCorr \cite{Blaber2015}. An example schematic of the queuing process for the \pyvale\ multi-threaded RG-DIC is illustrated in Fig. \ref{fig:rg_schematic}. In our method, we start with a single user specified seed location from which a single thread then calculates the correlation coefficient for this point and its four nearest neighbors. If the seed location and its neighbors successfully converge, then the neighboring points are added to a queue. This is a standard initial step in the RG-DIC algorithm. Where the algorithm of \pyvale\ differs is that other threads then look for \textit{work} by checking the queues of the other threads and \textit{stealing} the points from other threads if there are idle subsets waiting in the queue. Other threads always steal the point at the top of the other queues (highest correlation coefficient). This is then repeated until all threads have been allocated work in their own queue. Each thread then populates its own queue and continues along a thread specific RG-DIC path, checking and updating a global mask to prevent repeat calculations for individual subsets. Threads can run out of work - perhaps by hitting an ROI boundary or arriving at a point where all neighboring points have been calculated. In such cases, the thread will restart the process of searching for work in at the top of the queues of other active threads. If work is available, the thread will once again steal work and begin populating its own queue. This process repeats until correlation values have been computed across the entire ROI.
\section{Case Study: DIC Challenge 2.0}
\label{sec:DIC2}
\begin{figure*}[t!]
  \centering
  \includegraphics[width=1.0\linewidth]{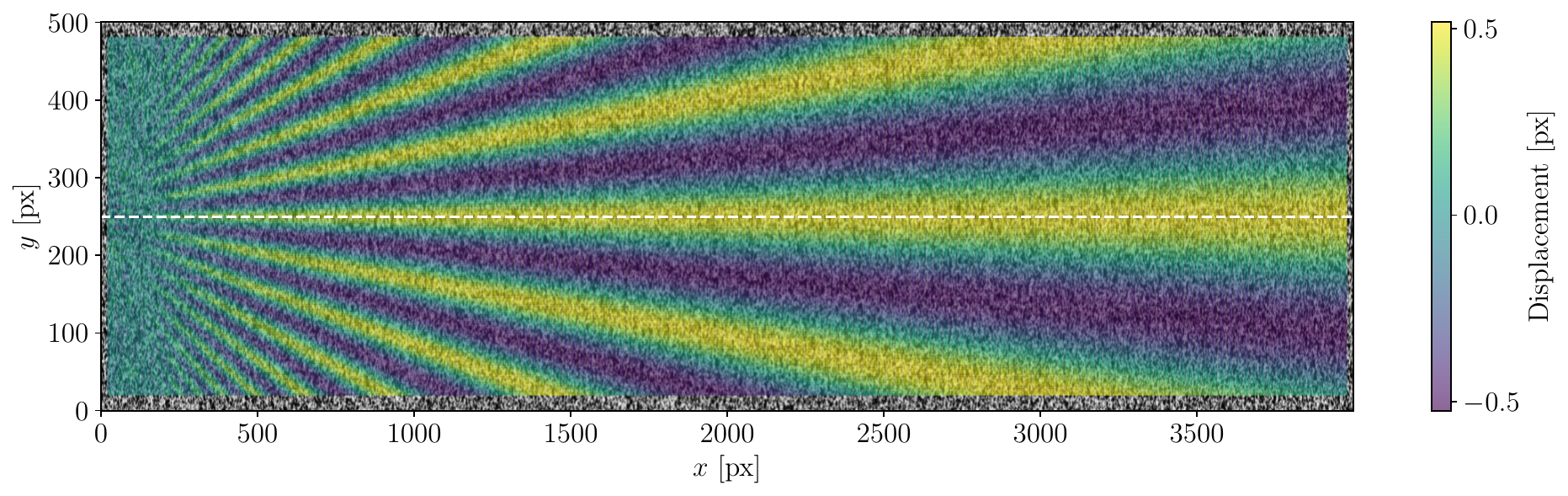}
  \caption{Example showing the displacement calculated using \pyvale's 2D DIC engine for the \textit{Star5} reference and deformed images from the DIC Challenge 2.0 dataset. The deformed image has a maximum ground truth displacement of $\pm 0.5$ pixels. The convergence of the DIC displacement data along the horizontal dotted white line at the vertical midpoint to the ground truth value of 0.5 px is used to determine metrological performance. Images can be downloaded from \url{https://idics.org/challenge/}.}
  \label{fig:star_pattern}
\end{figure*}
\begin{figure}[t!]
  \centering
  \includegraphics[width=1.0\linewidth]{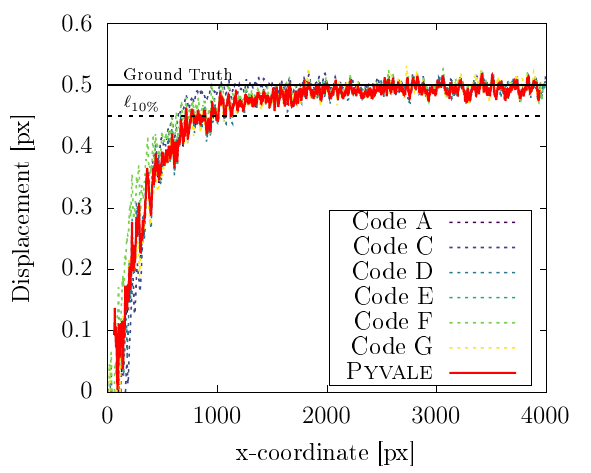}
  \caption{A example comparison of the convergence of \pyvale\ (Solid red line) and along the midpoint horizontal (shown by the dotted white line in Fig. \ref{fig:star_pattern}) anonymized DIC codes (dotted lines) to the ground truth displacement value of 0.5 px using the DIC challenge 2.0 images. }
  \label{fig:line_segment}
\end{figure}
\begin{figure}
  \centering
  \includegraphics[width=1.0\linewidth]{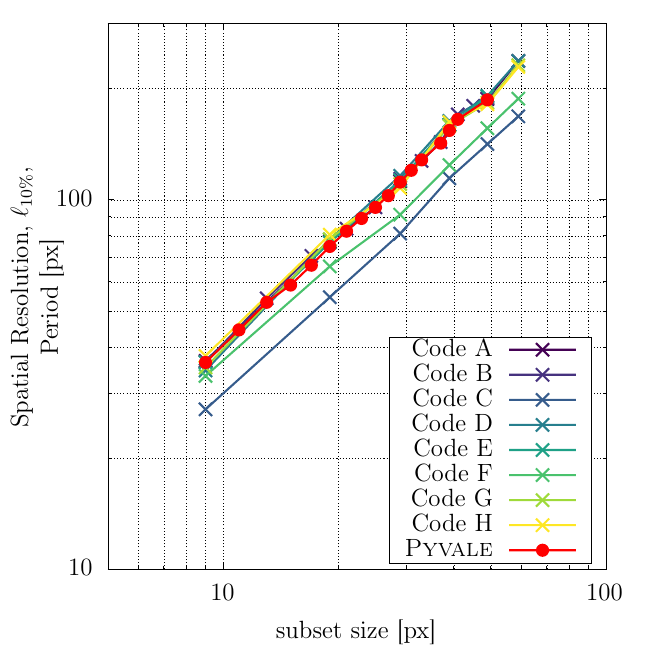}
  \caption{Spatial resolution of \pyvale\ (red circles) compared to other DIC codes (crosses) looked at as part of the DIC Challenge 2.0.}
  \label{fig:spatial_res}
\end{figure}

The Society of Experimental Mechanics (SEM) together with the International DIC (iDICs) society put together the DIC challenge to compare performance and accuracy of DIC codes to a set of well-defined standards (\url{https://idics.org/challenge/}). At the time of writing, there have been two iterations of the 2D DIC challenge \cite{dic_chal1, dic_chal2}, as well as stereo \cite{ahmad2024stereo} and Digital Volume Correlation (DVC) \cite{croom2021interlaboratory} comparisons. In this section, we shall compare the metrological performance of \pyvale\ to other local subset-based DIC codes that participated in the second iteration of the 2D DIC challenge. A short description of the benchmark images from the challenge are provided below. Those interested in a full in-depth description of the participants, methods and image generation procedures can find further details through Ref \cite{dic_chal2}.
\\
\\
The benchmark dataset created as part of the 2D DIC challenge 2.0 contains a series of images with a sinusoidal vertical displacement pattern with decreasing spatial frequency from left to right. In the DIC challenge paper \cite{dic_chal2}, these are reffered to as \textit{star} images. The same naming convention will be applied here. We present a comparison of \pyvale\ to other codes for \textit{Star5}. For this pattern, three reference images are provided: a noise free reference image, a reference image + noise, and a noisy deformed image. Each image is $4000 \times 500$ pixels with a maximum ground truth displacement of $\pm 0.5$ pixels at peak amplitude. An example of the star pattern as calculated from \pyvale's 2D DIC engine overlaid over the noisy reference image is shown in Fig. \ref{fig:star_pattern}.
\\
\\
The assessment of metrological performance in the DIC challenge stems from the ability of a DIC code to converge to the ground truth displacement of 0.5 pixels along the horizontal direction for a fixed vertical displacement at $y=250$ pixels (shown by the dotted white line in Fig. \ref{fig:star_pattern}). An example comparison of the convergence to the ground truth segment between \pyvale\ and the anonymized codes can be found in Fig. \ref{fig:line_segment}. This example calculation was performed for a subset-size and step of 19 and 1 pixels respectively using an affine shape function with a ZNSSD cost function. \pyvale\ is shown by the solid red line, while all other anonymized codes are shown with dotted lines. Note that not all codes provided results for this specific subset-size, hence the omission of \textit{Local.Affine.B} and \textit{Local.Affine.H}. While this example is only for a single subset-size, \pyvale's DIC engine can be seen to converge at an almost identical rate to other codes.
\\
\\
To fully assess the metrological performance of \pyvale\ we compare the spatial resolution, $\ell$, and Metrological Efficiency Indicator, $MEI$, to the anonymised results from different local subset DIC codes.  In the DIC challenge, an attenuation of 10\% was selected as a reasonable cutoff to define the spatial resolution, $\ell_{10\%}$. This cutoff is shown by the dashed horizontal line in Fig. \ref{fig:line_segment}. A measure of $\ell_{10\%}$ was obtained by fitting a 12th order polynomial to midpoint convergences to the ground truth displacement for varying subset-size. From the fit, the period of  oscillation (in pixels) at the $x$-coordinate where the fit crosses the 10\% bias threshold defines $\ell_{10\%}$ for a given subset-size. A subset-step of 1 was used for all subset-sizes. The period of oscillation at the 10\% cutoff for \pyvale\ alongside the anonymized codes from the DIC challenge can be found in Fig. \ref{fig:spatial_res}. While \pyvale\ agrees well with most codes, there are two that standout with a higher spatial resolution compared to the majority. As a result of the anonymization, we are unable to comment on the differences in implementations.
\\
\\
Using results for the spatial resolution, a comparison to the $MEI$ can be made. The MEI is defined by the measurement resolution multiplied by the spatial resolution \cite{grediac2017critical, dic_chal2}:
\begin{equation}
MEI = n \times \ell_{10\%}
\end{equation}
where $n$ is the displacement noise ($n=1\sigma$). This was obtained by performing a DIC calculation on the reference image against the reference image with added noise.
\begin{figure}[t!]
  \centering
  \includegraphics[width=1.0\linewidth]{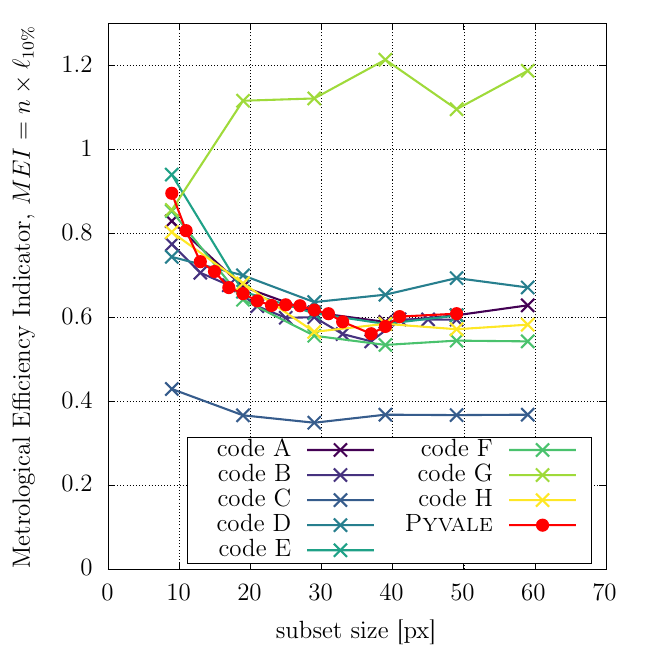}
  \caption{Spatial resolution of \pyvale\ (red circles) compared to other DIC codes (crosses) looked at as part of the DIC Challenge 2.0.}
  \label{fig:mei}
\end{figure}
Generally, the lower the MEI value, the better the metrological performance of a particular code. The MEI should theoretically remain constant as a function of subset-size, which has been analytically proven for Localized Spectrum Analysis (LSA) and empirically observed for both local and global DIC approaches\cite{grediac2017critical, blaysat2020towards}. Fig. \ref{fig:mei} shows a comparison between the MEI as a function of subset-size for \pyvale\ and the other DIC challenge codes. The final MEI value is calculated as the average of the three lowest values, which is inline with calculations in the DIC Challenge. A table of MEI values along with percentage differences to relative Code B can be found in Table \ref{tab:mei}.
\begin{table}[t!]
\centering
\begin{tabular}{lrr}
\toprule
\textbf{Code}   & \textbf{MEI}      & \textbf{\% Difference to Ref} \\
\midrule
A      & 0.602 & +6.38\%  \\
B      & 0.566 & \textbf{Ref} \\
C      & 0.361 & -36.20\% \\
D      & 0.654 & +15.54\% \\
E      & 0.600 & +5.95\%  \\
F      & 0.541 & -4.47\%  \\
G      & 1.023 & +80.68\% \\
H      & 0.573 & +1.30\%  \\
\midrule
\pyvale\ & 0.576 & +1.84\%  \\
\bottomrule
\end{tabular}
\caption{Average of the three lowest MEI values and percentage difference relative to code B (reference). Code B was used as a reference for the equivalent comparison in the DIC Challenge \cite{dic_chal2}.}
  \label{tab:mei}
\end{table}
This section has shown \pyvale\ has comparable metrological performance to other local-subset based DIC codes. Next we'll discuss  the capability to deal with extremely large SEM-DIC images (demonstrated in the next case study) and its computational efficiency compared to other open-source codes (demonstrated in the final case study).

\section{Case Study: Application of \pyvale\ to Experimental Scanning Electron Microscope DIC (SEM-DIC) images}
\label{sec:SEM}
\begin{figure*}[t!]
  \centering
  \includegraphics[width=1.0\linewidth]{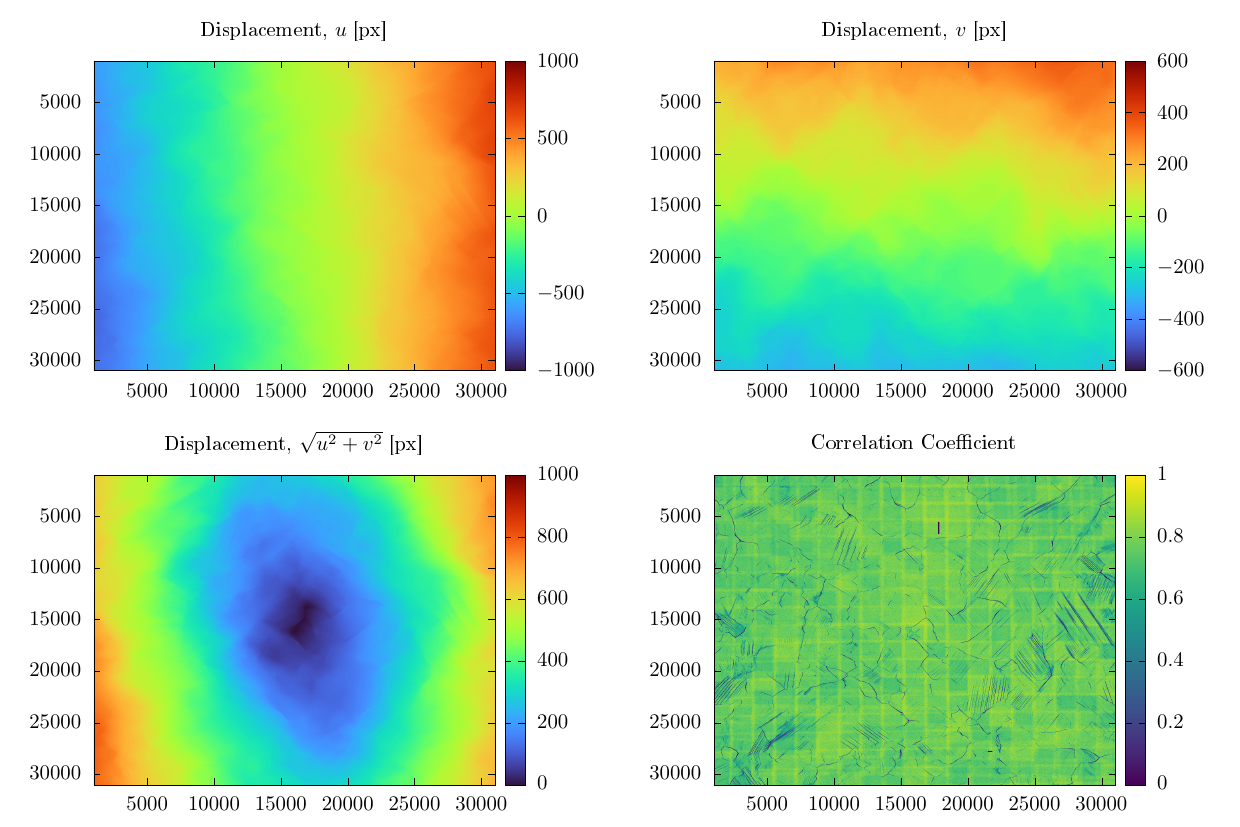}
  \caption{Displacement values and correlation coefficient output from \pyvale\ for an example $32000\times32000$ image using the FFT + RG algorithm.}
  \label{fig:sem_disp}
\end{figure*}

Here we will demonstrate \pyvale's ability to handle gigapixel-scale SEM-DIC images using a reference and deformed experimental image pair of $32,000\times32,000$ pixels. The experiment was performed on oxygen-free high conductivity copper and captured using a Tescan Clara field emission scanning electron microscope, operated at \SI{5}{\kilo\volt} accelerating voltage, with a \SI{300}{\pico\ampere} probe current and a \SI{3.2}{\micro\second} dwell time. To enable high resolution strain mapping over a large region of interest, a $19\times19$ array of \SI{20}{\micro\metre} $\times$ \SI{20}{\micro\metre}  tiles with a $2048\times2048$ image resolution were captured. A tile overlap of 20\% was used to ensure accurate image stitching following capture. Image arrays were captured automatically using the NewTec Scientific Softstrain software, controlling the SEM through the API interface. Images were stitched using the ITK-Montage Python package \cite{zukic2021itkmontage}. The tensile specimen was deformed to a global strain of approximately 5\% using a NewTec Scientific MT1000 microtensile tester. 
\\
\\
We used the \pyvale\ 2D DIC engine to analyse the SEM-DIC images with a subset-size and step of 31 and 15 pixels respectively, a ZNSSD correlation function, and an affine shape function with bicubic b-spline interpolation for sub-pixel accuracy. For each subset the maximum number of optimization iterations was set to 40 and convergence precision and convergence threshold set to 0.70 and 0.01 respectively. A Median Absolute Deviation (MAD) \cite{concise_mad_2008_mean} outlier removal was applied to the estimated rigid displacements from the multi-window FFT approach. This was done at each window size iteration. In many cases, MAD outlier removal will prevent incorrect peak locations in the FFT correlation data propagating down through the window sizes. The ROI was created programmatically, with a 1000 pixel boundary region excluded along all outer edges of the reference image. No subset correlations were performed in this border region. Output from DIC calculations were saved to a human-readable file in a standard comma-separated value (CSV) format. The maximum displacement was set to 2000 pixels, which is well above the expected maximum of $\sim 1000$ pixels. The horizontal, vertical and magnitude displacement maps, along with a correlation coefficient map can be found in Fig. \ref{fig:sem_disp}. The grid-like pattern that can be seen in the correlation coefficient map is a result of the stitching procedure used to combine the SEM data. Displacements have been saved for each and every subset in the correlation, hence the continuity in the displacement data. In many cases, a user may not want return displacement values for subsets that have not met the convergence criteria. In such a scenario a flag can be toggled to set all unconverged displacements to \texttt{NaN} values prior to saving to the results file. Further details can be found in the \pyvale\ documentation \cite{pyvale_docs}. 
\\
\\
\subsection{Runtime and Performance for SEM-DIC Images}
\begin{figure}[t!]
  \centering
  \includegraphics[width=1.0\linewidth]{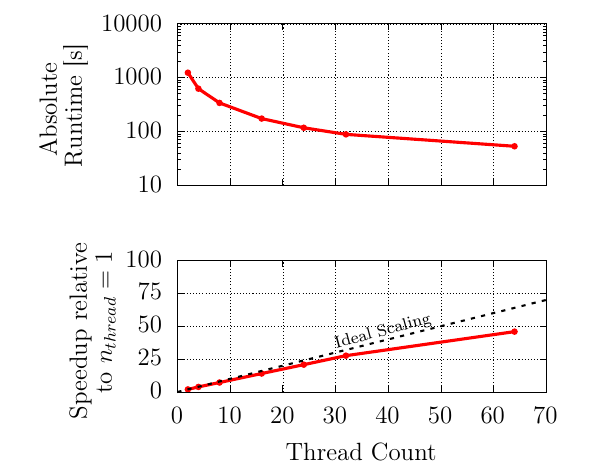}
  \caption{(Top) Absolute runtime and relative speedup (Bottom) for increasing thread count for the $32000 \times 32000$ pixel SEM image pair.}
  \label{fig:runtime}
\end{figure}

\begin{figure}[t!]
  \centering
  \includegraphics[width=1.0\linewidth]{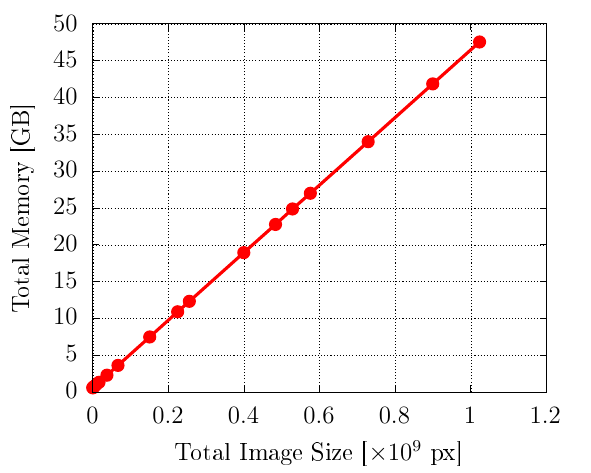}
  \caption{Peak memory consumption of \pyvale\ for varying image sizes. Thread count was fixed at 64.}
  \label{fig:scaling_memory}
\end{figure}

To illustrate how \pyvale\ scales in both memory usage and runtime for images of this size, we examine both runtime scaling as a function of thread count on a single workstation, and memory scaling with variable image sizes for a fixed number of threads. The workstation used for these tests had an AMD Ryzen Threadripper 7980X (64 cores), 256 GB of DDR5 RAM, 1TB NVMe SSD, running Ubuntu 24.04.2 LTS (x86\_64) as the operating system. Fig. \ref{fig:runtime} shows both the absolute runtime and the speedup relative to single-threaded performance for the SEM image pair. For the $32000\times32000$ pixel images analyzed here, processing with a single thread results in a correlation time of approximately 40 minutes. Using all 64 threads reduces this time to 48 seconds, nearly a 50-fold speedup. This is illustrated in the bottom panel of Fig. \ref{fig:runtime}. The dotted line in the bottom pane shows an ideal scaling relationship with a one-to-one relationship between the relative speedup and increasing thread count. 
\\
\\
In Fig. \ref{fig:runtime} \pyvale\ can be seen to drop below an ideal scaling relationship due to the growing amount of communication required between threads for the RG-DIC multi-threaded queuing. An alternative approach would be to have totally independent threads that work within an individual ROI (similar to Ncorr) where users must specify a seed location for each thread. This adds a significant amount of setup work for the user. Having to select all 128 seed locations and ensure convergence before running through the RG-DIC would be an arduous task. While increasing setup time, it is likely to lead to in an increase in scaling performance if we assume converged seed locations as well as image defects, poor lighting, artifacts, etc. are distributed evenly between the individual thread ROIs. However, it is possible to envisage scenarios where the ROI associated with an individual thread covers a mostly artifact ridden portion of the overall ROI causing a specific thread to be stuck. Such cases would lead to an inflated runtime, with threads finishing their ROI at different times. Threads in ``good'' regions of the ROI will sit idle while threads covering ``poor'' regions will require a greater number of optimization iterations per subset. 
\\
\\
Memory consumption and cache misses are important considerations for software performance, especially for the images used in this section. There are six main buffers in memory being responsible for most of the memory consumption in \pyvale's 2D DIC implementation. Namely, these are the reference image, the stack of deformed images, directional derivatives used for bicubic interpolation, and a boolean mask for the ROI. Fig. \ref{fig:scaling_memory} shows an almost perfectly linear scaling of \pyvale's memory usage as a function of image size for a single pair of reference and deformed images. The subset-size and subset-step were fixed at 21 and 10 pixels respectively. These quantities do have an impact on memory consumption, especially as subset-steps approach a single pixel. For the subset-step and subset-size used here, by far the most dominant quantity that affects memory usage is the image size. It is also worth noting here that thread count has little affect on memory consumption for standard problems, with thread local memory requirements mainly being dominated by subset information and optimization parameters, which are small compared to the shared image and interpolant buffers. 50 GB of memory was required in total.

\section{Comparison of Runtime to Other Open-source Software Packages}
\label{sec:runtime_comparison}
\begin{figure*}[t!]
  \centering
  \textbf{Synthetic Speckle Pattern}
  \includegraphics[width=1.0\linewidth]{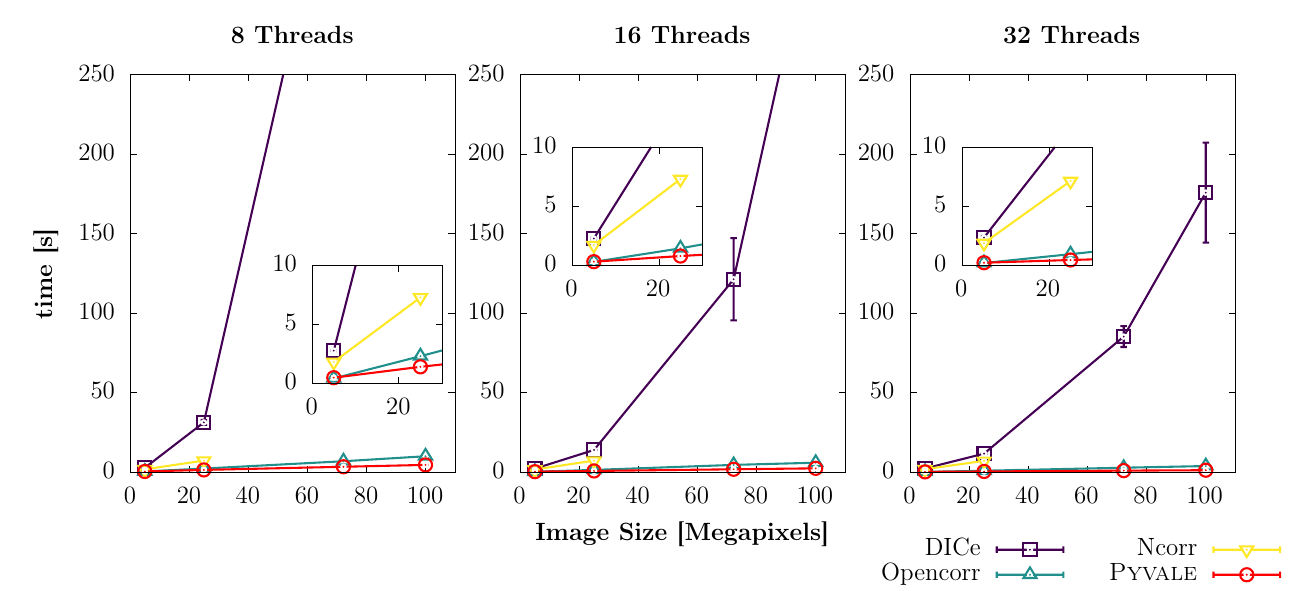}
  \textbf{SEM Images}
  \includegraphics[width=1.0\linewidth]{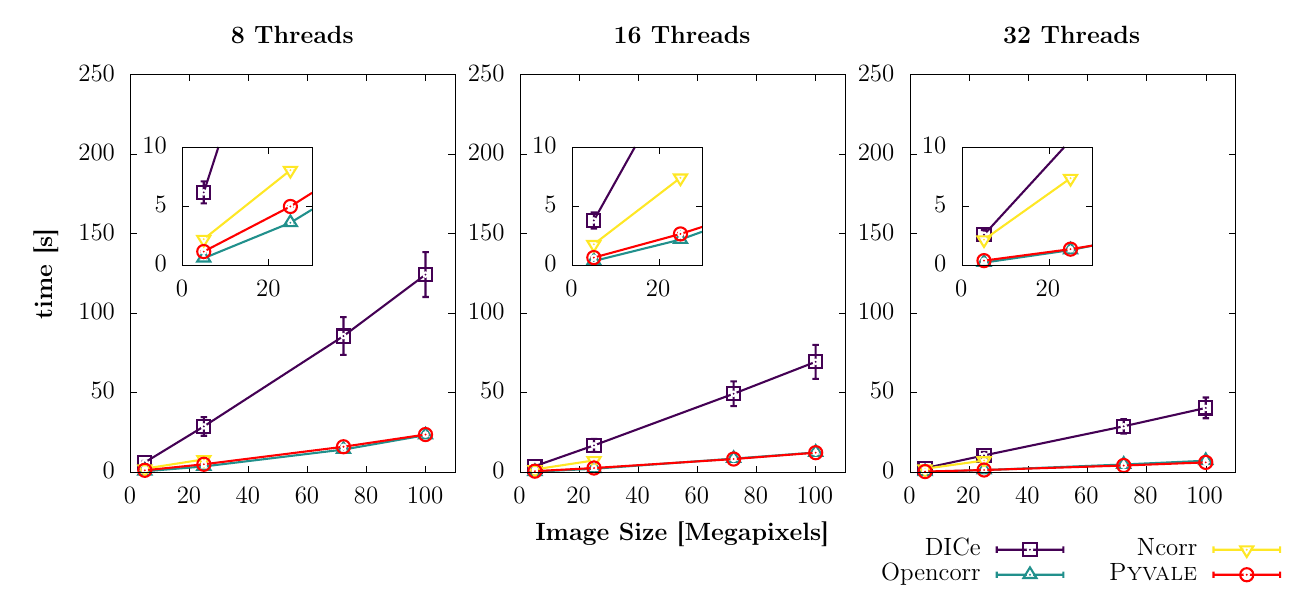}
  \caption{Correlation time for the the two image sets used in the open-source code comparison. The top row shows results for for the synthetically generated speckle pattern with a hydrostatic strain for varying image sizes and thread count. Bottom row shows runtime for the SEM images used in previous sections. Missing data points are due to programs failing due to exceeding RAM limit. Ncorr, OpenCorr and \pyvale\ all have similar correlation times and show a decrease in runtime for increasing thread count. DICe was shown to be considerably slower in for both image sets used here.}
  \label{fig:corr_comparison}
\end{figure*}

\pyvale\ was compared against three other open-source, local-subset-based DIC codes: OpenCorr \cite{jiang2023opencorr}, DICe \cite{Turner2015DICe}, and Ncorr \cite{Blaber2015}. All comparisons shown in this section were performed on the same workstation described in the previous section (AMD Ryzen Threadripper 7980X with 64 cores, 256 GB of RAM, running Ubuntu 24.04.2 LTS x86\_64). OpenCorr and DICe are implemented in C++, while Ncorr is a MATLAB extension that is heavily dependent on interactions within a GUI. There is some limited scripting functionality, but this is generally restricted to loading reference, deformed images and ROI selection. In contrast, both OpenCorr and DICe support extensive scripting and can be executed independently of a GUI, enabling a systematic study of runtime across varying image sizes and thread counts. Although the DIC codes examined here differ in documentation quality and maintenance status, each was installed and configured according to its respective documentation to ensure proper operation. Input parameters were aligned based on the available documentation wherever possible to facilitate a fair, like-for-like comparison. All scripts and results are openly available through \cite{pyvale_dataset}.
\\
\\
Two comparisons were conducted; The first used a synthetically generated speckle pattern stretched radially outward from the image center with a maximimum extension of 15 pixels along each edge. Image dimensions of $2,000\times2,500$, $5000\times5000$, $8,500\times8,500$, $10000\times10000$ pixels were used. Attempts with $32000\times32000$ were made, however all codes, except \pyvale, crashed due to exceeding the workstations amount of allocated RAM. The second comparison looked at performance under less ideal conditions involving larger displacements. Images of the same sizes to those listed above were generated from centered crops of the SEM dataset used in the previous section. A rectangular ROI was used with excluded boundaries of 100, 200, 300, and 400 pixels for the respective image dimensions. 
\\
\\
Correlation parameters were standardized across all codes where possible. The maximum subset optimization iteration count was set to 40, and the cutoff precision set to 0.01. The subset-size for \pyvale, OpenCorr, and DICe was fixed at 31 with a subset-step of 15. Ncorr, which uses circular subsets, employed a radius of 15 pixels (for a 30 pixel diameter). Thread counts of 8, 16, and 32 were tested. All C++ codes were compiled using GCC with the `-O3' optimization flag. Ncorr was executed using a standard MATLAB 2025b installation.
\\
\\
\begin{figure*}[t!]
  \centering
  \textbf{Synthetic Speckle Pattern}
  \includegraphics[width=1.0\linewidth]{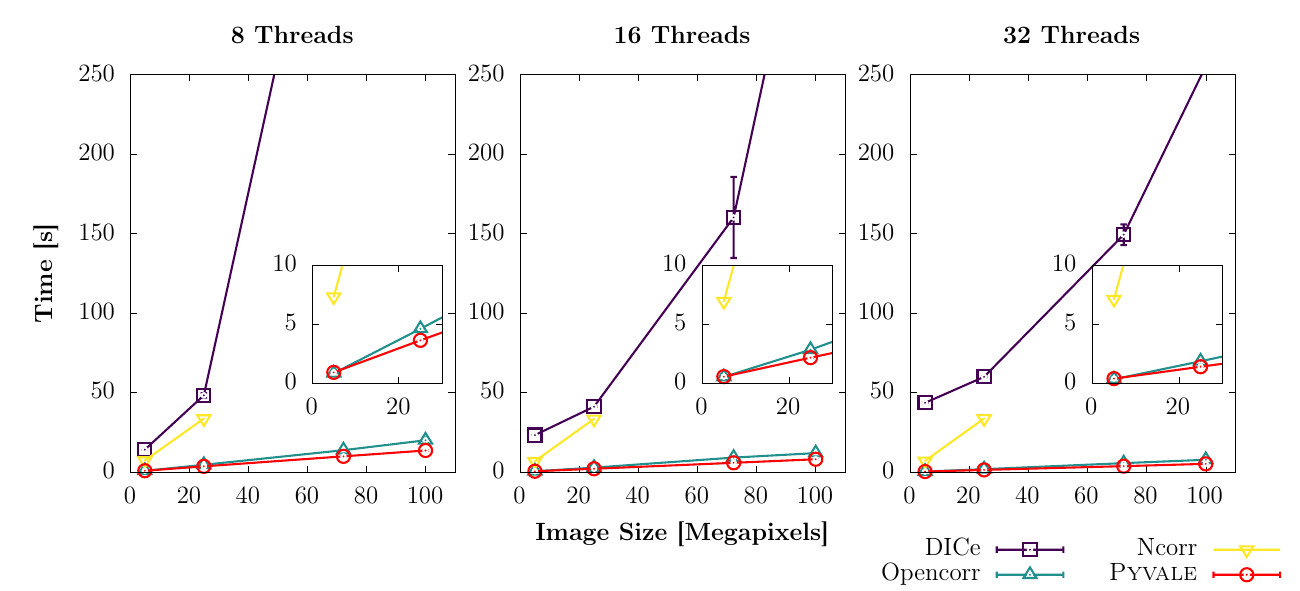}
  \textbf{SEM Images}
  \includegraphics[width=1.0\linewidth]{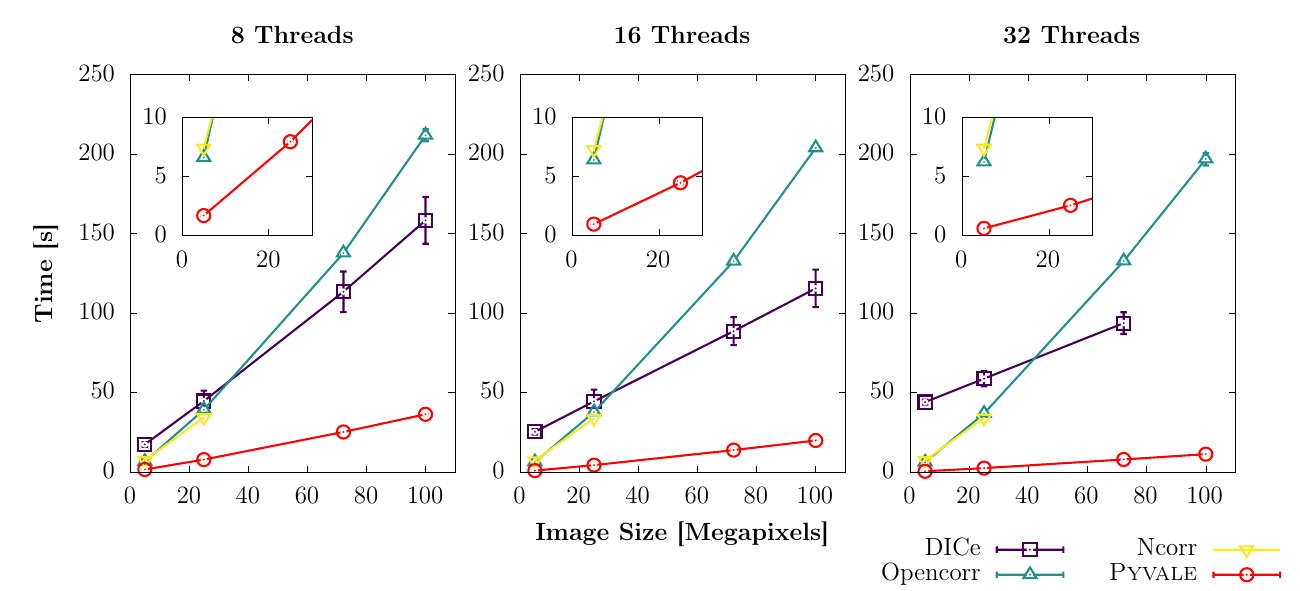}
  \caption{Total end-to-end runtime the two image sets used in the open-source code comparison. The top row shows results for the synthetically generated speckle pattern for varying image sizes and thread count. Bottom row shows runtime for the SEM images used in previous sections. Missing data points are due to programs failing due to exceeding RAM limit.}
  \label{fig:runtime_comparison}
\end{figure*}
The Ncorr MATLAB extension requires a seed location for the RG-DIC process in each thread; these were selected manually within the GUI with best efforts made to distribute them evenly across the ROI. OpenCorr’s initial displacement estimates were computed using a Scale-Invariant Feature Transform (SIFT) approach. Although OpenCorr also supports FFTCC initialization, its window size matches the subset-size used in the non-linear optimization, preventing accurate detection of displacements larger than the subset in a single pass. Implementing a multi-window FFTCC algorithm within the C++ setup script used by OpenCorr would have been possible but deemed to be beyond the expected scope for a typical user. Consequently, SIFT initialization was used for the large-deformation SEM images, while FFTCC initialization was applied to the synthetic speckle patterns which have smaller deformations. For DICe, a feature-matching initialization was used for both image sets.
\begin{figure*}[t!]
  \centering
  \textbf{Synthetic Speckle Pattern}
  \includegraphics[width=1.0\linewidth]{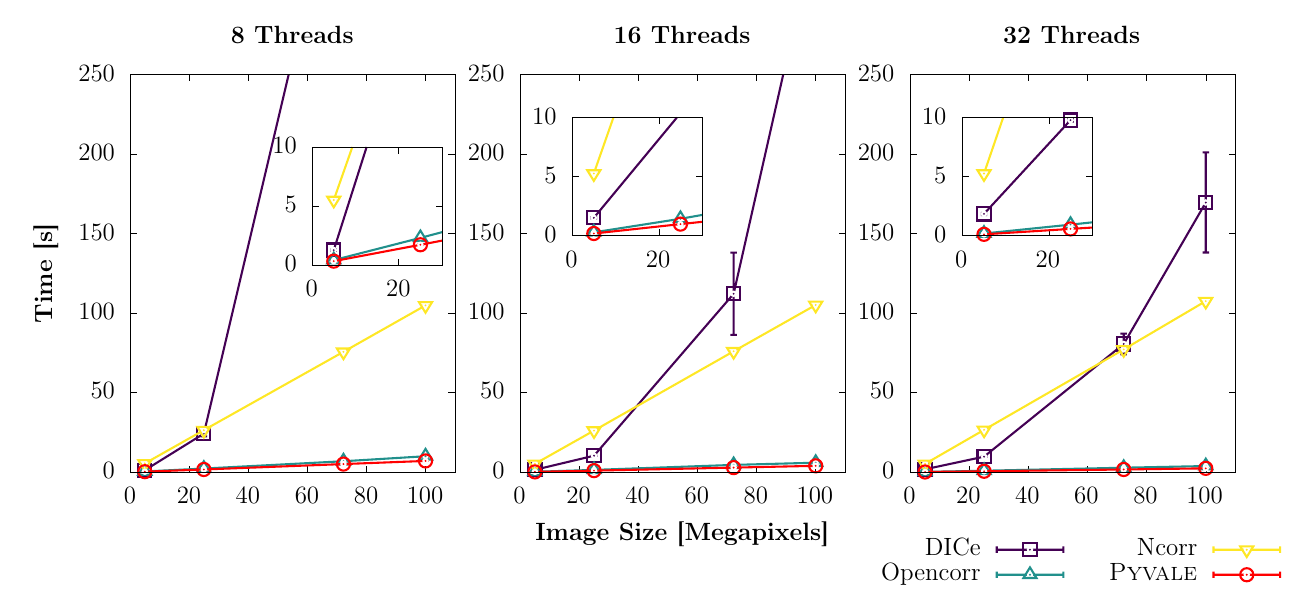}
  \textbf{SEM Images}
  \includegraphics[width=1.0\linewidth]{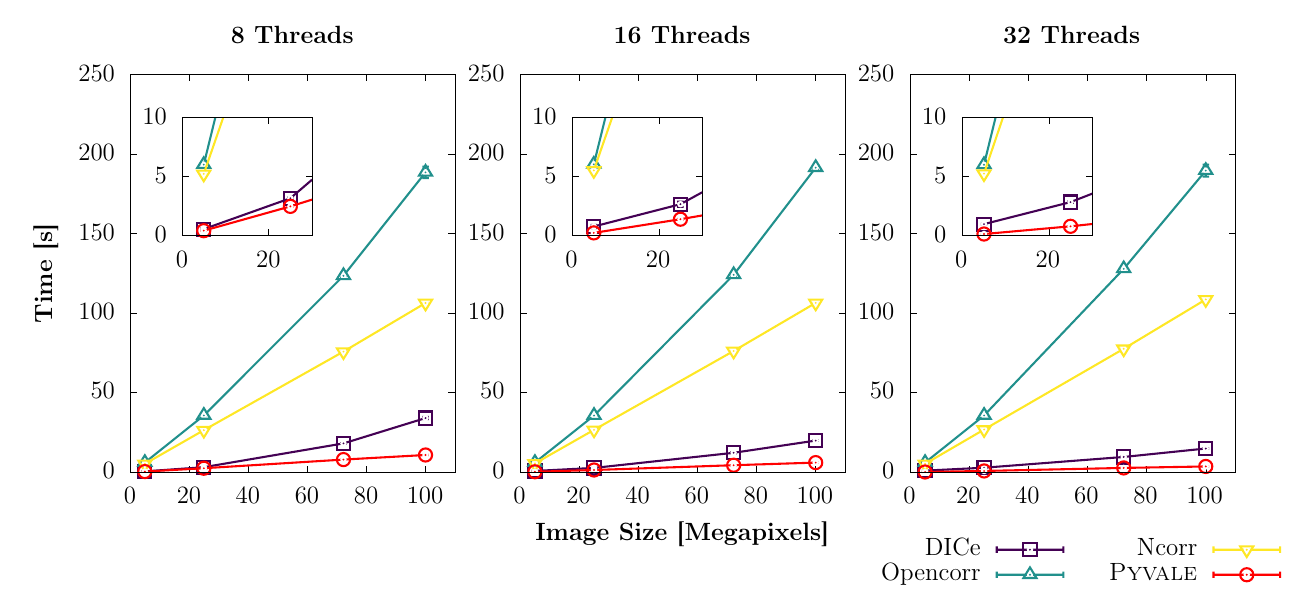}
  \caption{Initialization time for the two image sets used in the open-source code comparison. The top row shows results for for the synthetically generated speckle pattern with a hydrostatic strain for varying image sizes and thread count. Bottom row shows runtime for the SEM images used in previous sections. Missing data points are due to programs failing due to exceeding RAM limit. \pyvale\ and OpenCorr perform with similar speeds for the synthetic speckle pattern as both are using a FFTCC cross correlation approach. DICe's feature matching performs much better for the SEM images, likely due to the artifacts in the material providing suitable features to correlate between reference and deformed images. Ncorr shows consistent times across thread counts, it is therefore likely that the section of code responsible for the initialization does not make use of multi-threading.}
  \label{fig:init_comparison}
\end{figure*}
\\
\\
Presented runtimes are for full end-to-end execution performed for 3 repetitions for each code. For Ncorr, reported times exclude any GUI interactions and include only initialization and correlation. For DICe, OpenCorr, and \pyvale, runtimes are from script execution to termination. This naturally includes an initialization, correlation, terminal and file Input/Output (I/O). Runtime comparisons for the three thread counts on both the synthetic speckle and SEM datasets are shown in Fig. \ref{fig:runtime_comparison}. Missing data points indicate cases where a code has crashed across all three repeats due to running out of available memory. \pyvale\ and OpenCorr show comparable performance for the synthetic speckle images with runtime decreasing as thread count increases. In these tests, both codes use an FFTCC on a single subset sized ($31\times31$) window to estimate rigid displacements before performing non-linear optimization to compute displacement fields. Ncorr’s runtime remains nearly constant across all thread counts. The long initialization time (relative to the correlation time) does not change significantly with increasing thread count, in contrast to the correlation time, which decreases as more threads are used. A breakdown into initialization and correlation runtimes is provided across Fig. \ref{fig:init_comparison} and Fig. \ref{fig:corr_comparison}.
\\
\\
Another noticeable difference between datasets is the increased runtime of OpenCorr for SEM images relative to synthetic patterns. This is attributed to the initialization approach: their FFTCC initialization efficiently captures small deformations (within a single subset) in synthetic images, whereas the SIFT-based approach required for large SEM deformations introduces much greater computational overhead compared to the multi-window FFT approach used in \pyvale.
\\
\\
Along with runtime performance, installation time and setup complexity are also important considerations for typical DIC users. For \pyvale, the goal is to make the installation process as straightforward as possible. By a considerable margin, Python remains the most widely used programming language for engineering and scientific applications \cite{Cass2025_TopProgrammingLanguages, TIOBE_Index}. As long as the user has a compatible version of Python and \texttt{pip} installed, \pyvale\ can be installed with a single command: \texttt{pip install pyvale}. This eliminates the need to manually manage commonly used C++ dependencies for image processing, such as OpenCV \cite{opencv_library} and FFTW \cite{fftw}, or work with build systems such as CMake \cite{cmake_reference_doc}.
\section{Future Development}
\label{sec:future}
Specifically for the 2D DIC module, several enhancements are planned. One is the integration of an automated reference-image updating scheme \cite{Bing2012}, in which the reference image evolves incrementally through the sequence of deformed images to accommodate large strains. Another planned improvement is an automated system for partitioning large images into smaller sequential chunks that can be loaded into memory one at a time during processing. This will allow \pyvale\ to handle arbitrarily large images efficiently, even on machines with typical amounts of memory (e.g., 8–32 GB of RAM). While \pyvale's DIC engine is currently limited to 2D, a key milestone on the development roadmap is the implementation of stereo DIC with full stereo calibration capabilities. Another major objective is to extend \pyvale\ to benefit from GPUs, especially to accelerate the multi-window FFT initialization utilizing the availability of specific FFT libraries \cite{gpufft1, gpufft2, gpufft3} for this hardware. Several existing DIC codes have already demonstrated substantial performance gains using this type of hardware \cite{gpu1,gpu2,gpu3}. Alongside continued development of the DIC engine itself, efforts will also focus on integrating the engine into the broader planned \pyvale\ ecosystem. This includes coupling the DIC engine with synthetic image deformation driven by FEA simulations, as well as enabling experimental design investigations through optimization of sensor placement and lighting using ray-tracing and rasterization graphics engines.
\\
\\
Beyond the planned new features listed above, ongoing work will focus on improving usability of the existing code base.  Requests for new features, examples, or bug reports can be submitted through the \pyvale\ GitHub repository \cite{pyvale_github}.
\section{Summary}
\pyvale\ has been designed to be an open-source and easy-to-use package for DIC with a simple installation procedure. By creating a simple Python interface that hooks into faster lower level languages for computationally heavy tasks, \pyvale\ is developing into a tool that is both accessible to new users and capable of handling the increasingly large images and datasets without compromising on either scalability or speed. Here, we have demonstrated the capabilities of \pyvale\ through:
(i) a numerical validation against the DIC challenge 2.0, 
(ii) the application of \pyvale\ to gigapixel resolution SEM-DIC images, with displacements approaching 1000's of pixels.
(iii) a performance comparison to other local-subset based open source codes demonstrating \pyvale\ is the most performant code for large SEM-DIC images.

While metrological performance is comparable for other local-subset DIC codes, what sets \pyvale\ (and global DIC codes such as muDIC \cite{olufsen2020mudic}) apart is the free open-source, easy-to-use Python interface. This, coupled with strong numerical performance will hopefully allow \pyvale\ to become a highly utilized code within the experimental mechanics community.
\\
\\
\pyvale\ is a free and open-source package, and contributions from the DIC community are encouraged, including for commercial use (\pyvale\ is distributed under the MIT license). Users are welcome to explore, modify, and extend the software to suit their own research needs, and to deploy as many copies of \pyvale\ as they see fit, whether on individual workstations and laptops for small-scale testing or on high-performance computing clusters for systematic sweeps of DIC parameters. Looking ahead, we aim for this project to support ongoing community-driven development and to provide a solid platform for future improvements in open-source DIC.

\section{Authorship Contribution Statement}
\begin{itemize}[itemsep=0pt]
\item\textbf{Joel Hirst:} Methodology, software writing, software validation, Article writing - original draft, Article writing - review \& editing.
\item\textbf{Lloyd Fletcher:} Conceptualization, software writing, software validation, article writing - review \& editing, project administration, funding acquisition.
\item\textbf{Adel Tayeb:} Conceptualization, software validation, project administration.
\item\textbf{Lorna Sibson, Megan Sampson, Wiera Bielajewa:} Software writing, software validation.
\item\textbf{Ben Poole:} Article writing - original draft, Experimental SEM work, software validation.
\item\textbf{Michael Atkinson, Alex Marsh, Rory Spencer, Rob Hamill:} Software validation, 
\item\textbf{Cory Hamlin, Allan Harte:} Funding acquisition, project administration.
\end{itemize}

\section{Acknowledgments}
We acknowledge support from UKRI through the Future Leaders Fellowship scheme, grant MR/Y015916/1. We also acknowledge support from the UK Engineering and Physical Sciences Research Council (EPSRC), grant EP/W006839/1.

\section{Data Availability Statement}
The source code for \pyvale\ can be accessed through the Github repository \cite{pyvale_github}. All scripts and results for the comparisons made in this work can be found in \cite{pyvale_dataset}.

\bibliographystyle{elsarticle-num} 
\bibliography{ref.bib}

\end{document}